\documentstyle[prd,aps,preprint]{revtex}

\def\lsim{\\lower-1.5pt\vbox{\hbox{\rlap{$<$}\lower5.3pt\vbox{\hbox{$\sim$}}}}\ }
\def\gsim{\\lower-1.5pt\vbox{\hbox{\rlap{$>$}\lower5.3pt\vbox{\hbox{$\sim$}}}}\ }

\newcommand{\be}{\begin{equation}}
\newcommand{\ee}{\end{equation}}

\def\bq{\begin{eqnarray}}
\def\eq{\end{eqnarray}}

\def\th{\theta}

\begin{document}
\tightenlines

\title{Brane corresponding to the Nariai bulk}

\author{Naresh Dadhich$^{a}$ and Yuri Shtanov$^{b}$}

\address{$^a$Inter-University Centre for Astronomy and Astrophysics,
Post Bag 4, Ganeshkhind, Pune~411~007, India. \\ $^b$Bogolyubov Institute for
Theoretical Physics, Kiev 03143, Ukraine}

\maketitle

\begin{abstract}
We consider the five-dimensional bulk spacetime with negative $\Lambda$
described by the Nariai metric (which is not conformally flat) and match it
with a vacuum brane satisfying the proper boundary conditions.  It is shown
that the brane metric corresponds to a cloud of string dust of constant energy
density.
\end{abstract}

\bigskip


\bigskip

A fresh impetus to the old paradigm of spacetime with extra dimensions was
recently given in \cite{dvali}, where it was suggested that compact extra
dimensions may be macroscopic while our space-time is described as a
lower-dimensional domain wall (brane) where all the matter is concentrated. A
novel approach to higher-dimensional braneworld cosmology emerged after Randall
and Sundrum postulated the existence of a {\em noncompact\/} spacelike fifth
dimension \cite{RS}. According to this world-view, our perception of `normal'
four-dimensional gravity arises because we live on a domain wall (brane)
embedded in or bounding a `bulk' anti-de~Sitter space (AdS)\@. The metric
describing the full (4+1)-dimensional space-time is non-factorizable, and the
small value of the true five-dimensional Planck mass is related to its large
effective four-dimensional value by the extremely large warp of the
five-dimensional space. The novelty of the Randall--Sundrum (RS) model is to
use the curvature of the bulk spacetime (with the $Z_2$ symmetry of reflection
relative to the brane) to keep zero-mass gravitons localized on the brane.

This theory was studied in detail in the case of 5D anti de~Sitter (AdS) bulk
with flat or Schwarzschild vacuum brane and in the cosmological context. The
bulk and brane solutions are matched by the Israel boundary conditions. The
effective Einstein equation on the brane can be written \cite{SMS} by using the
Gauss--Codazzi relations. It would additionally involve square of the
stress-energy tensor and projection of the bulk Weyl curvature tensor to the
brane. The latter is trace-free and is known as the Weyl dark energy/radiation.
In this sense, the system of equations on the brane is obviously not closed. It
is therefore very difficult to find exact complete solutions with both bulk and
brane metrics satisfying the proper boundary conditions. There exist only a few
examples of complete solutions, among which the AdS bulk with flat or
Schwarzschild brane and Schwarzschild--AdS bulk with FRW brane. Most of other
solutions including black hole \cite{DMPZ} and collapse \cite{BGM,GD} are
solutions of only the brane equations without the corresponding solution in the
bulk.

The purpose of this paper is to give one more simple example of complete
solution.  Specifically, the Nariai metric \cite{Nariai} offers an interesting
case of the Einstein space which is not conformally flat.  After the
generalization of this metric to 5D case with negative $\Lambda$, the question
of graviton confinement was studied for this conformally non-flat bulk
spacetime \cite{param}, and it was shown that there exist no normalized modes
for massless graviton (once again, there is a pointer to fine tuning of
parameters inherent in the Randall--Sundrum (RS) model \cite{PS,DK}). However,
this was done with the bulk metric alone without any reference to the brane
spacetime. In this paper, we complete the solution by finding the corresponding
brane satisfying the proper boundary conditions.

The theory that we consider here is described by the following general action
(see \cite{CHS}):
\begin{equation} \label{action}
S = M^3 \sum \left[\int_{\rm bulk} \left( {\cal R} - 2 \Lambda \right) - 2
\int_{\rm brane} K \right] + \int_{\rm brane} \left( m^2 R - 2 \sigma \right) +
\int_{\rm brane} L \left( h_{\alpha\beta}, \phi \right)
\end{equation}
in the standard notation, where the sum is taken over the bulk components
bounded by the brane.  We use the signature and sign conventions of
\cite{Wald}.  The lagrangian $L \left( h_{\alpha\beta}, \phi \right)$
corresponds to the presence of matter fields $\phi$ on the brane and describes
their dynamics, and the extrinsic curvature $K_{\alpha\beta}$ of the brane is
defined with respect to the inner normal $n^a$, as it is done in
\cite{Shtanov}. Note that we have included the curvature term in the action for
the brane which arises when one incorporates quantum effects generated by
matter fields residing on the brane.

The equations in the bulk and on the brane are obtained from the variation of
Eq.~(\ref{action}), which gives
\begin{equation} \label{bulk}
{\cal G}_{ab} + \Lambda g_{ab} = 0 \, ,
\end{equation}
\begin{equation} \label{brane}
m^2 G_{\alpha\beta} + \sigma h_{\alpha\beta} = \tau_{\alpha\beta} + M^3 \sum
\left(K_{\alpha\beta} - h_{\alpha\beta} K \right) \, ,
\end{equation}
where $h_{\alpha\beta}$ is the induced metric on the brane,
$\tau_{\alpha\beta}$ is the stress-energy tensor resulting from the Lagrangian
$L \left( h_{\alpha\beta}, \phi \right)$, and the sum of the extrinsic
curvatures on either side of the brane is taken.

The Nariai metric in the bulk as given in Ref.~\cite{param} reads as
\begin{equation} \label{nariai}
 d s^2_5 = e^{- 2k|y|} \left( - dt^2 + dr^2 \right) + dy^2 + \frac{1}{2 k^2}
\left(d \th^2 + \sinh^2 \th d \phi^2 \right) \, .
\end{equation}
It is a solution of the bulk equation (\ref{bulk}) with $\Lambda = -3k^2$.
Since the Weyl curvature is non-zero for this metric and hence its projection,
$E_{\mu \nu} := C_{\mu a \nu b} \, n^{a} \, n^{b}$ on the brane
$y = \mbox{const}$, would be non-zero.

Now we consider the brane located at $y = 0$ which has induced metric
$h_{\alpha\beta}$ given by the line element
\begin{equation}
ds^2_4 = - dt^2 + dr^2 + \frac{1}{2 k^2} \left(d \th^2 + \sinh^2 \th d \phi^2
\right) \, .
\end{equation}
This is a spacetime having the structure of the product of a flat 2-dimensional
space and a 2-sphere of constant curvature \cite{nd}. Further, it can be shown
that the stress-energy tensor corresponding to this metric describes a cloud of
string dust \cite{let,bose}.

The stress-energy tensor for a string-dust distribution is given by
\cite{let,bose},
\begin{equation}
T_{\rm string}^{\mu\nu} = \rho \Sigma^{\mu\beta} \Sigma^{\nu}_{\beta} \, ,
\end{equation}
where $\rho$ is the proper energy density of the cloud, and $\Sigma^{\mu \nu}$
is the bivector associated with this world-sheet: $\Sigma^{\mu\nu} =
\displaystyle \epsilon^{AB} {\partial x^\mu\over \partial \xi^A} {\partial
x^\nu\over
\partial \xi^B}$. Here, $\epsilon^{AB}$ is the 2D Levi-Civita tensor (normalized
so that $\epsilon^{AB} \epsilon_{AB} = 2$) and $\xi^A = (\xi^0, \xi^1)$ are the
coordinates on the string world-sheet. Following Refs.~\cite{let,bose}, we
readily conclude that the stress-energy tensor corresponding to the above brane
metric accord with the string-dust stress-energy tensor, which satisfies the
equation of state $T^0{}_0 + T^i{}_i = 0$, a typical of topological defects
like cosmic string and global monopole.

The components of the Einstein tensor of this metric in the coordinates $(t, r,
\theta, \phi)$, are given by
\begin{equation}
G^\alpha{}_\beta = {\rm diag} \left( 2k^2, 2k^2, 0, 0 \right) \, .
\end{equation}
Clearly, the above-mentioned equation of state for the string-dust distribution
is satisfied and the string dust has constant negative energy density $\rho =
-2k^2$.

The extrinsic curvature on either side of the brane is given by
\begin{equation}
K^\alpha{}_\beta = {\rm diag} ( -k, -k, 0, 0 ) \, , \quad K = - 2k \, .
\end{equation}
Substituting it into Eq.~(\ref{brane}), we obtain the system of equations for
the vacuum brane $\left( \tau_{\alpha\beta} = 0 \right)$
\begin{equation}
2 m^2 k^2 + \sigma = 2 M^3 k \, , \quad \sigma = 4 M^3 k \, ,
\end{equation}
whence we get the conditions of ``fine tuning''
\begin{equation} \label{fine}
M^3 = - m^2 k < 0 \, , \quad \sigma = - 4 m^2 k^2 < 0 \, .
\end{equation}

The brane metric thus describes a cloud of string dust of constant negative
energy density. Furthermore, we see that the solution (\ref{nariai}) with brane
located at $y =0$ requires both the five-dimensional Planck mass $M$ and brane
tension $\sigma$ also to be {\em negative\/}.  According to the fine-tuning
conditions (\ref{fine}), only two of the four fundamental constants in this
theory are independent; for example, the constants $M$ and $\sigma$ can be
expressed in terms of $m$ and $\Lambda$ via Eq.~(\ref{fine}). In contrast to
the original RS model, our solution requires $m \ne 0$, i.e., it requires the
presence of the induced curvature term in the action for the brane.

The Nariai bulk and the corresponding string-dust brane may not appear to be of
much cosmological and astrophysical importance.  However, they undoubtedly
represent an interesting spacetime solution, and the Nariai metric has already
seen a good bit of application in the context of black hole and quantum-gravity
considerations \cite{odin}. Our aim was, in view of the paucity of complete
exact solutions for the bulk-brane system, to present one more example
involving very simple spacetimes.

{\it Acknowledgments:\/} Yu.~S. would like to thank IUCAA for warm hospitality
and acknowledges partial support from the INTAS grant for project No.~2000-334.

\end{document}